\documentclass[
reprint,
twocolumn,
superscriptaddress,
amsmath,amssymb,
aps,
pre
]{revtex4-1}

\usepackage{graphicx}% Include figure files
\usepackage{color}

\begin{document}
%\title{Giant diffusion in a noisy tilted periodic potential}
\title{Colossal Brownian yet non-Gaussian diffusion induced by nonequilibrium noise}
\author{K. Bia{\l}as}
\affiliation{Institute of Physics, University of Silesia, 41-500 Chorz{\'o}w, Poland}
\author{J. {\L}uczka}
\affiliation{Institute of Physics, University of Silesia, 41-500 Chorz{\'o}w, Poland}
\author{P. H\"anggi}
\affiliation{Institute of Physics, University of Augsburg, D-86135 Augsburg, Germany}
\author{J. Spiechowicz}
\email{jakub.spiechowicz@us.edu.pl}
\affiliation{Institute of Physics, University of Silesia, 41-500 Chorz{\'o}w, Poland}

\begin{abstract}
We report on novel Brownian, yet non-Gaussian diffusion, in which the mean square displacement of the  particle grows linearly with time, the probability density for the particle spreading is Gaussian-like, however, the probability density for its  position increments possesses an exponentially decaying tail. In contrast to recent works in this area, this behaviour is not a consequence of either a space or time-dependent diffusivity, but is induced by external non-thermal noise acting on the particle dwelling in a periodic potential. The existence of the exponential tail in the increment statistics leads to colossal enhancement of diffusion, surpassing drastically the previously researched situation known under the label of ``giant'' diffusion. This colossal diffusion enhancement  crucially impacts  a broad spectrum of the first arrival problems, such as diffusion limited reactions governing transport in living cells.
\end{abstract}

\maketitle
\section{Introduction}
Brownian diffusion is in the limelight of present activities and enjoys seemingly never-ending interest  \cite{kanazawa2020, neupane2016, jeon2016, spiechowicz2016njp, peng2016, kim2017, zhang2017, illien2018, goychuk2019, spiechowicz2019njp} across a broad spectrum of scientific disciplines, extending even into socio-economics where  diffusion of ideas and innovations has been considered recently \cite{buera2020, rogers2003}. It presents an archetype stochastic processes which is characterized by two fundamental features. The first is its proportionality of the mean square displacement $\sigma_x^2(t)$ to elapsed time, namely,
\begin{equation}
\sigma_x^2(t) = 2 D t,
\end{equation}
where $D$ is the diffusion coefficient. The second is rooted in the Gaussian shape of the probability density function (PDF) to observe the entity at position $x$ at time $t$, i.e.,
\begin{equation}
p(x,t) = \frac{1}{\sqrt{4 \pi D t}} \exp{\left(-\frac{x^2}{4 D t} \right)}.
\end{equation}
%The increment of the Brownian diffusion is of the Gaussian form as well.
The universal emergence of the Gaussian statistics is attributed to the central limit theorem which constitutes a cornerstone result for statistical physics \cite{feller1970}.

Recently, a new class of diffusion processes has been reported in a growing number of systems, which typically are of  biological origin, such as  soft and active matter setups \cite{wang2012}. In the latter dynamics the mean square displacement $\sigma_x^2(t)$ exhibits the linear growth in time, however, the corresponding PDF is distinctly non-Gaussian and in cases attains an exponential decay. %sometimes called a Laplace distribution.
Such an exponential behaviour is generally valid for transport in random media \cite{jl1,jl2,barkai2020}, as e.g. for glassy systems \cite{shell2005}. This Brownian, yet non-Gaussian diffusion has been explained so far by the classical idea of superstatistics \cite{wang2009,hapca2009} and by other approaches assuming a diffusing diffusivity \cite{chubynsky2014,jain2016,metzler2017}.

In this work we demonstrate yet a new class of Brownian dynamics, being non-Gaussian diffusion in which, however, the mean square displacement $\sigma_x^2(t)$ is still a linear function of elapsed time and the PDF $P(x, t)$ to observe the entity at position $x$ at time $t$ is very close to Gaussian; but the corresponding PDF $p(\Delta x)$ for the increments of the process is distinctly non-Gaussian, exhibiting a non-conventional exponential tail. The latter fact is in clear contrast to usual Brownian diffusion for which the increments are distributed as well according to a Gaussian PDF. Particularly, such non-Gaussian behaviour is induced here by a stochastic, impulse-like external forcing on the system.  This is different from  previous approaches where anomalous features were a consequence of either space- or time-dependent diffusion coefficients, reflecting the characteristic features of the particle environment \cite{metzler2017}. Last but not least, the existence of the exponential tail in the statistics of increments leads to truly colossal enhancement of diffusion.
\begin{figure*}[t]
    \includegraphics[width=0.45\linewidth]{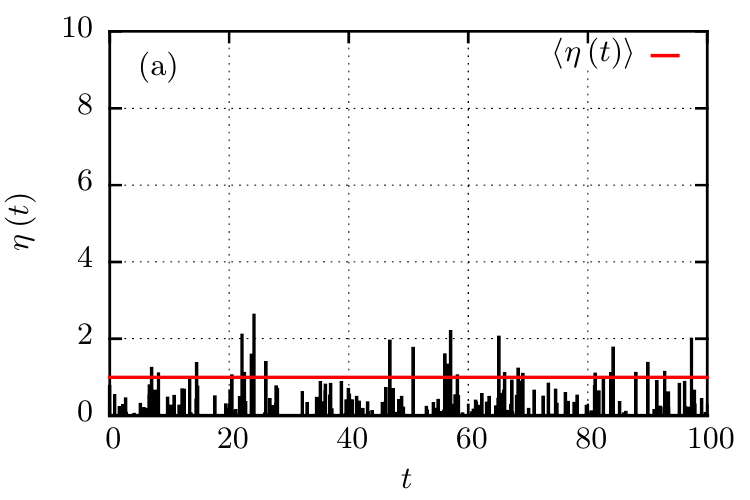}
    \includegraphics[width=0.45\linewidth]{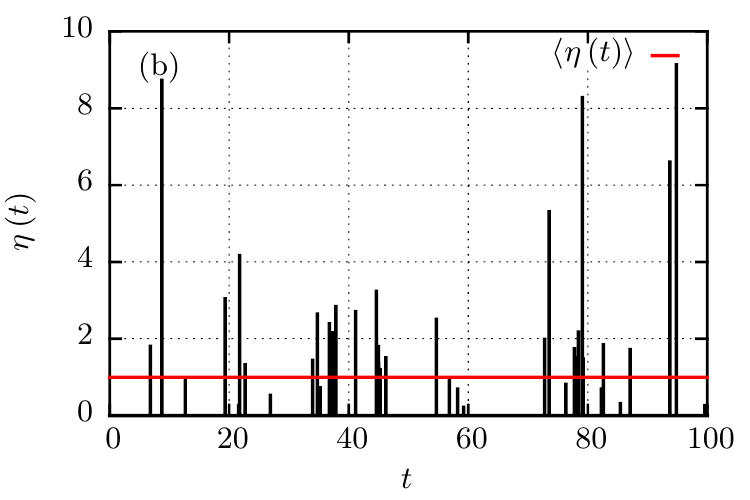}
    \caption{(color online) Exemplary realizations of Poisson noise for different mean spiking rate $\lambda$ as well as intensity $D_P$ for the fixed average value $\langle \eta(t) \rangle = 1$, as indicated by  the (red) solid line. Panel (a): $\lambda=2$, $D_P=0.5$; (b) $\lambda=0.5$, $D_P=2$.}
    \label{supp_fig1}
\end{figure*}

For this purpose we consider a variant of an archetypal model for a nonequilibrium system, namely, an overdamped dynamics of a Brownian particle diffusing in a periodic  potential $U(x) = \sin{x}$ under the action of a static tilting force $f$ \cite{risken}. It came as a surprise for the community that the diffusion coefficient for such a system at a critical deterministic bias $f \sim f_c$, near threshold towards running deterministic solutions, pronouncedly exceeds the free diffusion value, thus leading to the phenomenon known as giant diffusion \cite{constantini1999,reimann2001a,reimann2002,lindner2001,lindner2016,spiechowicz2020pre}. This type of giant diffusion is theoretically well understood \cite{reimann2002} and corroborated experimentally \cite{lee2006,reimann2008,ma2015}. We analyse the variant of the above system by replacing the static force $f$ with a biased nonequilibrium noise $\eta(t)$ which pumps energy to the system in a random way.  Such a case is of paramount interest in understanding of transport properties of not only physical but also biological systems, e.g. living cells \cite{bressloff2013}, where for a strongly fluctuating environment there is no systematic deterministic load but rather random collisions or releases of chemical energy which here are modeled in the form of kicks and impulses. In order to be able to make a comparison with the deterministic bias $f$ we set the mean value of the noise force $\eta(t)$ equal to $f$, namely $\langle \eta(t) \rangle = f$. As we will show below, the nonequilibrium noise $\eta(t)$ may drive the particle in such a way that the diffusion coefficient grows arbitrarily large above  the studied giant diffusion case with a deterministic bias force $f$. This biased noise is shown to be at the source for the non-Gaussianity of the PDF for the increments of particle positions.
\section{Model}
Let us start with the overdamped Langevin dynamics for position of the Brownian particle, which in  dimensionless variables reads
\begin{equation}
	\label{model_eta}
    \dot{x}=-U'(x)+\sqrt{2 D_T}\,\xi(t) +\eta(t).
\end{equation}
We refer the readers to Appendix wherein we detail on the scaling procedure. The dot and prime denote differentiation with respect to time $t$ and coordinate $x$ of the Brownian particle, respectively. %The potential $U(x)$ is assumed to be in the {\it symmetric} form \mbox{$U(x)=\sin{x}$}.
Thermal fluctuations are modelled by $\delta$-correlated Gaussian white noise $\xi(t)$ of vanishing mean $\langle \xi(t) \rangle = 0$ and the  correlation function $\langle \xi(t)\xi(s) \rangle = \delta(t-s)$. Its intensity $D_T$  is proportional to temperature $T$ of ambient thermal bath, i.e. $D_T \propto T$ (see Eq. (A.6) in Appendix).

As an example of the stochastic biasing force $\eta(t)$ we propose a sequence of $\delta$-shaped pulses with random amplitudes $z_i$ defined in terms of biased white Poisson shot noise (PSN) \cite{spiechowicz2013jstatmech, hanggi1978, hanggi1980}, i.e.,
\begin{equation}
    \eta(t)=\sum^{n(t)}_{i=1}z_i\delta (t-t_i),
\end{equation}
where $t_i$ are the arrival times of a Poissonian counting process $n(t)$, characterized by the parameter $\lambda$; i.e. the PDF for occurrence of $k$ impulses in the time interval $[0,t]$ is given by the Poisson probabilities \cite{feller1970}
\begin{equation}
	Pr\{ n(t) = k \} = \frac{(\lambda t)^k}{k!} e^{-\lambda t}.
\end{equation}
The parameter $\lambda$ can be interpreted as the mean number of $\delta$-pulses per unit time. The amplitudes $\left\{ z_i \right\}$ are independent random variables  distributed according to a common PDF $\rho\left(z\right)$. The latter distribution is assumed to be of an exponential form, i.e. $\rho(z) = \zeta^{-1}\, \theta(z)\, \mbox{exp}(-z/\zeta)$, where the parameter $\zeta >0$ and $\theta(z)$ denotes the Heaviside step function. As a consequence, all amplitudes $\{z_i\}$ are positive of mean value
 $\langle z_i \rangle = \zeta$ and realizations of the force are \emph{non-negative}, i.e.,  $\eta(t) \ge 0$. This presents white noise of finite mean and a covariance given by \cite{spiechowicz2014pre}
\begin{subequations}
\begin{align}
	\langle \eta(t) \rangle &= \lambda \langle z_i \rangle = \sqrt{\lambda D_P}, \\
	\langle \eta(t)\eta(s) \rangle - \langle \eta(t) \rangle \langle \eta(s) \rangle &= 2 D_P \delta (t-s),
\end{align}
\end{subequations}
where we introduced the PSN intensity $D_P = \lambda \langle z_i^2 \rangle/2 = \lambda \zeta^2$. We also assume that thermal fluctuations $\xi(t)$ are uncorrelated with nonequilibrium noise $\eta(t)$, i.e. $\langle \xi(t) \eta(s) \rangle = \langle \xi(t) \rangle \langle \eta(s) \rangle = 0$. The impact of PSN parameters $\lambda$ and $D_P$ on its stochastic realizations is presented in Fig. \ref{supp_fig1}. There we depict two example trajectories for different mean spiking rates $\lambda$ as well as noise intensities $D_P$ for a fixed average value $\langle \eta(t) \rangle = 1$. As it can be deduced from these two panels, the parameter $\lambda$ may be interpreted as the frequency of the $\delta-$spikes whereas $D_P$ is proportional to the amplitude of the single pulse. We mention that e.g. if simultaneously $\lambda$ is large and $D_P$ small, then the particle is frequently kicked by small impulses. On the other hand, if $\lambda$ is small and $D_P$ is large then it is rarely kicked by large spikes. It is worth to note that in the limiting case $\lambda \to \infty$, $\zeta \to 0$ with $D_P = \lambda \zeta^2 = const.$ PSN tends to Gaussian white noise of  intensity $D_P$.

The Markovian stochastic dynamics given by Eq. (\ref{model_eta}) yields for the probability density $P(x,t)$ of the process $x(t)$ the integro-differential master equation \cite{hanggi1978,hanggi1980}
\begin{eqnarray}
\label{continuity}
\frac{\partial}{\partial t} P(x,t) = \frac{\partial}{\partial x}
[U'(x) P(x,t)] + D_T \frac{\partial^2}{\partial x^2} P(x,t) \nonumber \\ + \lambda \int_{-\infty}^{\infty} [P(x-z,t) -P(x,t)] \rho(z) \,dz.
\end{eqnarray}
%where the probability current $J(x,t)$ reads
%\begin{align}
%\label{current}
%J(x,t) &= [-U'(x) + f - \lambda \langle z_i \rangle] P(x,t) - D_T \frac{\partial}{\partial x} P(x,t) \nonumber\\ &+ \lambda \int_{-\infty}^{\infty} \rho(z) \int_0^z P(x-y,t) \,dy dz.
%\end{align}
%The master Eq. (\ref{continuity}) together with Eq. (\ref{current}) can be interpreted as a  spatially nonlocal diffusion equation with an effective diffusion function consisting of nonlocal (Poissonian) and local (thermal) parts \cite{czernik1997}.

The observable of foremost interest for this study is the diffusion coefficient, being  defined as
\begin{equation}
	D= \lim_{t \to \infty} \frac{\sigma_x^2(t)}{2t} = \lim_{t \to \infty} \frac{\langle x^2(t) \rangle - \langle x(t) \rangle^2}{2t}\;,
\end{equation}
where $\sigma_x^2(t)$ is the variance of the particle position $x(t)$ and  the average value reads
\begin{equation}
	\langle x^k(t) \rangle = \int_{-\infty}^{\infty}  x^k P(x,t) dx.
\end{equation}
The diffusion coefficient $D$ for the overdamped dynamics obeying Eq. (\ref{model_eta}) with the deterministic constant force $f$  has been calculated in  a closed analytical form in \cite{reimann2001a,reimann2002}. %The expression in the dimensional units reads
%\begin{equation}
%    \label{analytical}
%    D=D_0 \frac{\int_{x_0}^{x_0+L}\frac{dx}{L}I_{+}^2\left(x\right)I_{-}\left(x\right)}{\left[\int_{x_0}^{x_0+L}\frac{dx}{L}I_{+}\left(x\right)\right]^3},
%\end{equation}
%here $D_0$ is the Einstein's free diffusion coefficient \mbox{$D_0:=k_B T/\Gamma$} ($\Gamma$ stands for the friction coefficient), $x_0$ is the arbitrary reference point and $I_{\pm}(x)$ is defined as
%\begin{equation}
%    I_{\pm}\left(x\right):=\int_{0}^{L}\frac{dy}{D_0}e^{\left\{\pm U\left(x\right)\mp U\left(x\mp y\right)-yF\right\}/k_B T}.
%\end{equation}
Those authors detected that for weak thermal noise and near the critical tilt $f_c = 1$, the diffusion may become greatly enhanced as compared to free diffusion. In such a case the corresponding washboard potential $V(x)=U(x)-fx$ exhibits a strictly monotonic behaviour and exactly one inflection point within each period. The dynamics for weak thermal fluctuations is mainly determined by two processes, namely (i) noise induced escape from the potential minimum and (ii) relaxation towards the next minimum \cite{risken}. The relaxation time is robust with respect to thermal noise intensity whereas the escape time exhibits an exponential sensitivity. This dichotomous microdynamics can be detected from the particle trajectories as depicted in Fig. \ref{supp_fig3} (a). %For realistic experimental setup it leads to enhancement of free diffusion of up to 14 orders of magnitude so that in principle thermal spreading of particles should be observable on a macroscopic scale at room temperature \todo{cite}.

In this work we demonstrate how the latter effect of giant enhancement of diffusion is affected when the static force $f$ is replaced by stochastic forcing $\eta(t)$ in the form of biased PSN. In this case the dynamics is described by the  integro-differential master equation (\ref{continuity}) and its solution $P(x, t)$  can no longer be expressed in a closed analytical form.  Therefore, we investigate this problem by means of the precise numerical simulations. In this approach, the averaging is over the initial conditions $x(0)$  distributed uniformly over
the spatial  period $[0, 2\pi]$ of the potential $U(x)$ as well as  over Gaussian $\xi(t)$ and PSN $\eta(t)$ noise realizations. For details of the latter we refer the readers to Ref. \cite{spiechowicz2015cpc}.
\begin{figure}[t]
	\centering
    \includegraphics[width=0.9\linewidth]{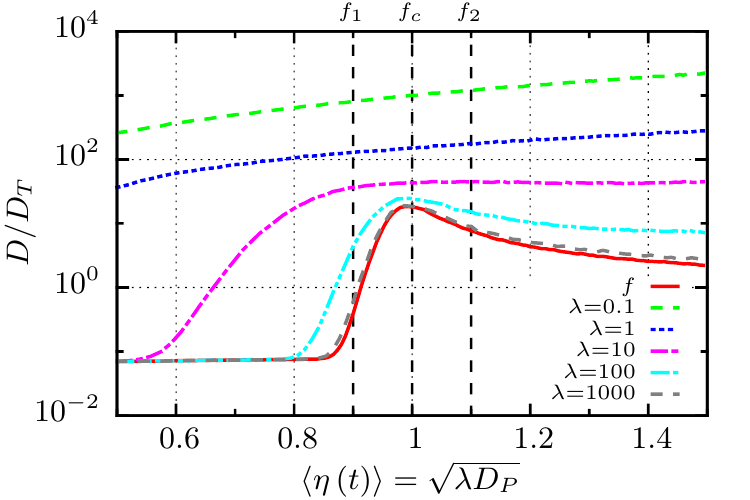}
    \caption{(color online) The relative diffusion coefficient $D/D_T$, where $D_T$ corresponds to free thermal diffusion, {\it vs.} the average value $\langle \eta(t) \rangle = \sqrt{\lambda D_P} = f$ of the mean external force in the form of the biased PSN. This characteristic is depicted for different magnitudes of the spiking rate $\lambda$ and fixed temperature,  $D_T = 0.01$. The red solid line represents the diffusive behaviour driven by the deterministic tilting force $f$.}
    \label{fig1}
\end{figure}
%\vspace{-12pt}
\section{Results}
In Fig. \ref{fig1} we present dependence of the relative diffusion coefficient $D/D_T$
% where $D_T$ is the value corresponding to free thermal diffusion,
on the average $\langle \eta(t) \rangle$ of the biased PSN $\eta(t)$ for different magnitudes of the spiking rate $\lambda$. The (red) solid line corresponds to the diffusive behaviour of the particle under the action of the corresponding static tilting force $f$. The reader can observe therein the  known effect of diffusion enhancement, being   most pronounced near the critical tilt $f = f_c = 1$, i.e. when deterministic running solutions set in. The effect of biased stochastic forcing $\eta(t)$ is dual. First, PSN enhances much more strongly the diffusion coefficient $D/D_T$, see also Fig. \ref{fig3} (a). %Second, the interval for which the latter quantity is magnified over the free thermal diffusion is significantly broadened when the particle is driven by PSN.
Second, when the particle is rarely kicked by  large $\delta$-pulses, i.e. for $\lambda \to 0$ and $D_P \to \infty$ with $\langle \eta(t) \rangle = \sqrt{\lambda D_P} = const.$, the maximum in the relative diffusion coefficient $D/D_T$ near the critical force $f = f_c = 1$ disappears indicating that the particle motion is decoupled from the periodic potential. This is expected because in this limit the PSN force  dominates  a contribution  of  $U(x)$.

\begin{figure*}[t]
	\centering
	\includegraphics[width=0.45\linewidth]{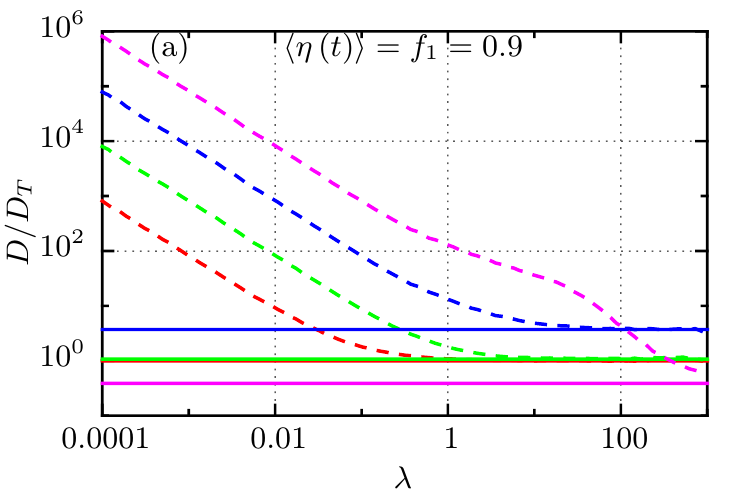}
	\includegraphics[width=0.45\linewidth]{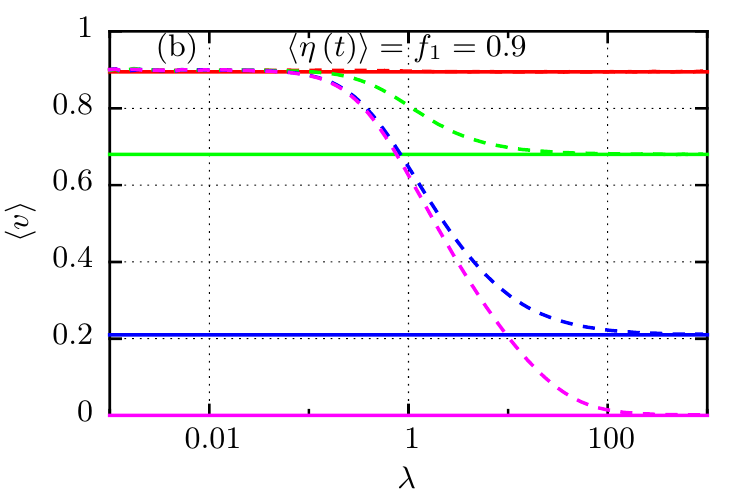}\\
	\includegraphics[width=0.45\linewidth]{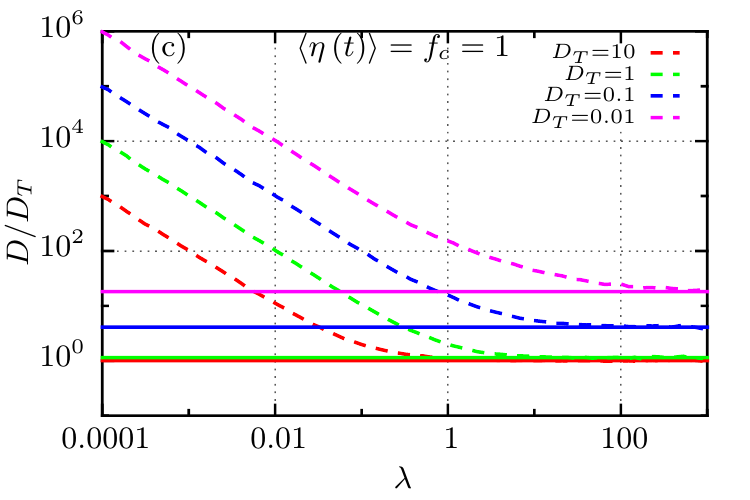}
	\includegraphics[width=0.45\linewidth]{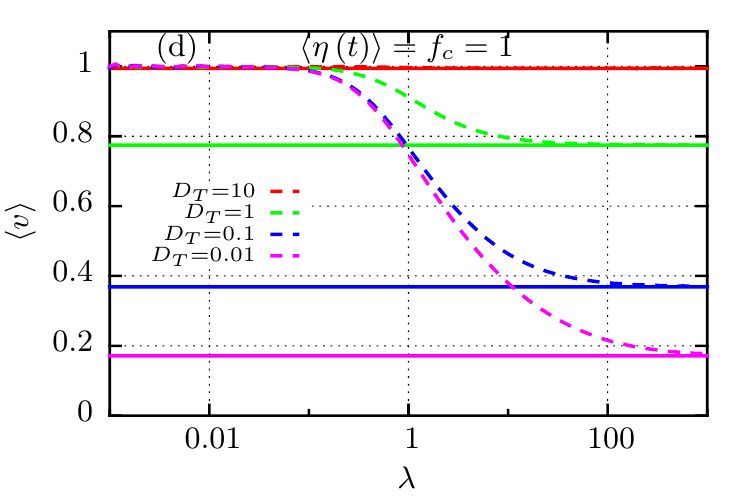}\\
	\includegraphics[width=0.45\linewidth]{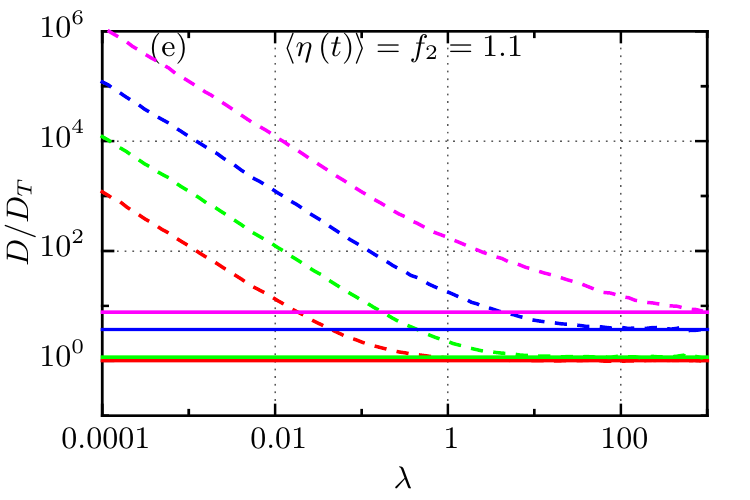}
	\includegraphics[width=0.45\linewidth]{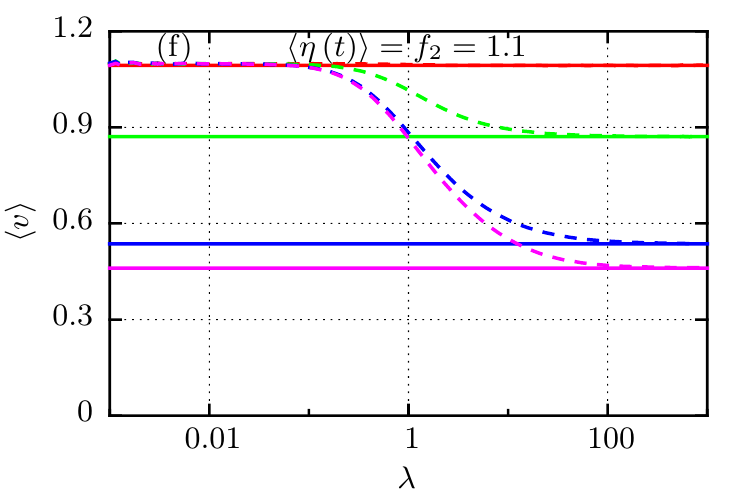}	
	\caption{(color online) Left column: the relative diffusion coefficient $D/D_T$, right column: the average velocity $\langle v \rangle$ of the particle, all presented as a function of the spiking rate $\lambda$ depicted for selected temperatures of the system,  $D_T \propto T$. The solid straight lines correspond to the above quantities  for the system with the static force $f$, whereas dashed ones indicates the influence of the PSN $\eta(t)$. Upper row (panels (a) and (b)): $\langle \eta(t) \rangle = f_1 = 0.9$, middle row (panels (c) and (d)): $\langle \eta(t) \rangle = f_c = 1$, bottom row (panels (e) and (f)): $\langle \eta(t) \rangle = f_2 = 1.1$, c.f. Fig. \ref{fig1}.}
	\label{fig2}
\end{figure*}

In Fig. \ref{fig2}  we depict the diffusion coefficient $D/D_T$ as a function of the spiking frequency $\lambda$ for selected values of $\langle \eta(t) \rangle$. The impact of different temperatures is also displayed.  The solid straight lines represent the diffusion coefficient for the corresponding system with the static bias $f$, whereas the dashed ones depict the influence of PSN. One  notices that  magnitude of the diffusion coefficient $D/D_T$ in the case of PSN is equivalent to the corresponding one for the deterministic force $f$ in the limiting case of large $\lambda$ and small $D_P$, i.e. for very frequent $\delta$-kicks of tiny amplitudes \cite{spiechowicz2014pre}. On the contrary, for rarely occurring, very strong random kicks $\lambda \to 0$ and $D_P \to \infty$, the relative diffusion coefficient $D/D_T$ is \emph{divergent}. It is an instructive example:  for the thermal noise intensity $D_T = 0.01$ and the system subjected to the critical tilt $f = 1$, the effective diffusion coefficient $D/D_T = 18$, meaning that it is \emph{18 times greater} than thermal diffusion $D_T$ for the free Brownian particle. For the same $D_T=0.01$, when  the particle is driven by Poissonian noise $\langle \eta(t) \rangle = f = \sqrt{\lambda D_P} = 1$ with $\lambda = 1$ and $D_P = 1$, i.e. the mean value of Poissonian noise amplitude $\langle z_i \rangle = \zeta = 1$ (half of the rescaled potential $U(x)$ barrier), the relative diffusion coefficient $D/D_T = 151$. This implies that it is \emph{nearly one order of magnitude greater} than for the already giant diffusion observed when the particle is subjected to the corresponding constant bias. Therefore, to emphasize this fact, we term it colossal diffusion. Moreover, as  $D_T$ decreases (i.e. temperature decreases) the enhancement of diffusion $D/D_T$ over the value observed for the static bias $f$ starts to be detected for the progressively larger spiking rates $\lambda$, or with $f_c=1=\lambda \langle z_i \rangle$ correspondingly smaller amplitudes $z_i$ of the $\delta$-kicks.
\begin{figure*}[t]
	\centering
	\includegraphics[width=0.45\linewidth]{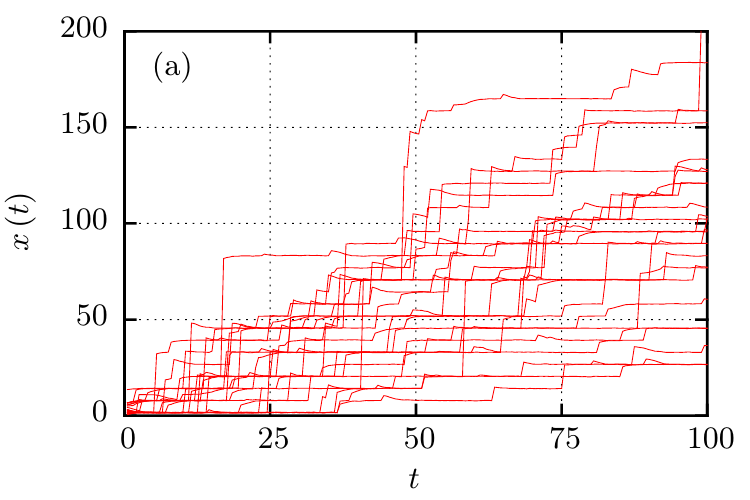}
	\includegraphics[width=0.45\linewidth]{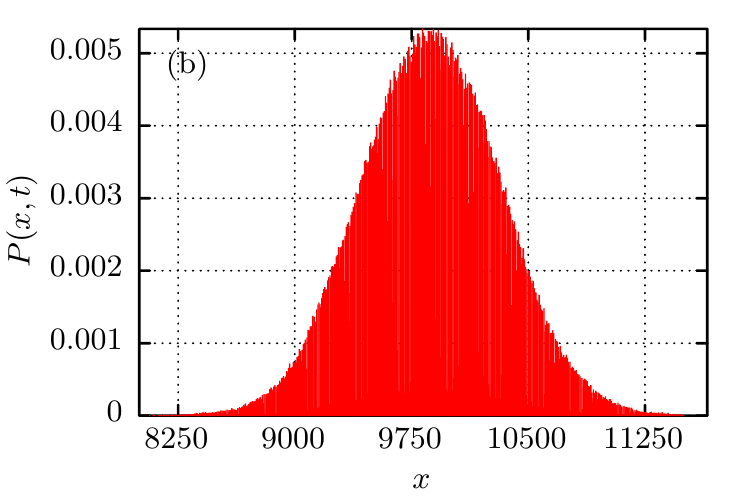}\\
	\includegraphics[width=0.45\linewidth]{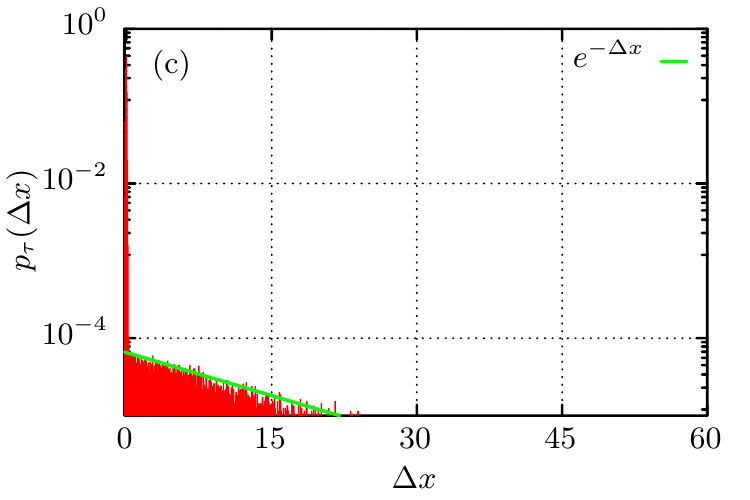}
	\includegraphics[width=0.45\linewidth]{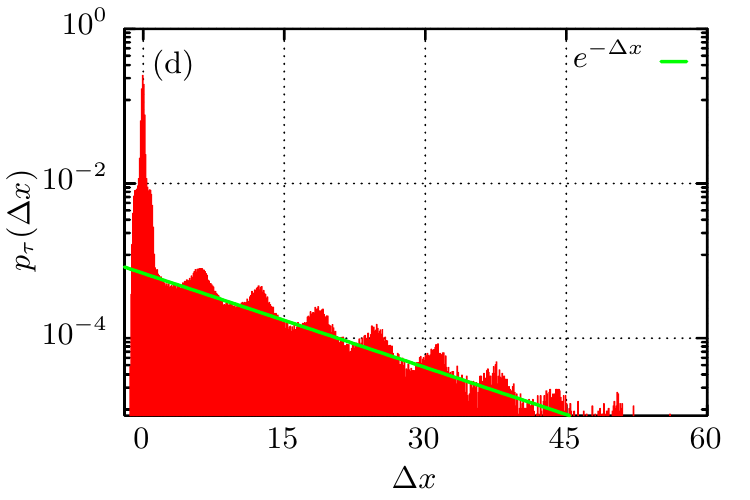}
	\caption{(color online) Overdamped Brownian particle driven by the biased PSN $\eta(t)$. Panel (a): An exemplary set of realizations of stochastic  system trajectories. Panel (b): The PDF $P(x,t)$ for the particle coordinate $x$ at time $t$.  Panels  (c) and (d): The PDF $p_{\tau}(\Delta x)$  for the long time particle position increments $\Delta x(\tau) = \lim_{t \to \infty}[x(t + \tau) - x(t)]$  is depicted for the time difference $\tau = 0.1$ in panel (c) and for $\tau = 1$ in panel (d).  Other parameters are: the thermal noise intensity  $D_T = 0.01$, the spiking rate $\lambda = 0.1$ and the Poisson noise intensity $D_P = 10$ (i.e., $\langle \eta(t) \rangle =1$). The above PDFs were calculated for $t = 10000$ for which we checked that $\sigma_x^2(t) \sim D t$. The exponential fits are indicated with the (green) lines.}
	\label{fig3}
\end{figure*}

In Fig. \ref{fig2}   we additionally depict the average velocity $\langle v \rangle = \lim_{t \to \infty} (x(t) - x(0))/t$ of the particle {\it vs} the mean frequency $\lambda$ of $\delta$-spikes. This transport quantifier is likewise notably enhanced when the static bias $f$ is replaced with $\eta(t)$. %The amplification is particularly impressive in the low temperature regime where it may be of orders of magnitude.
The main difference is that $\langle v \rangle$ does not diverge when the spiking rate tends to zero $\lambda \to 0$ but for $\lambda < 0.1$ it saturates at the value corresponding to the free diffusion $\langle v \rangle = f/\gamma = 1$ (recall that in the dimensionless units the friction coefficient $\gamma = 1$ and in the considered parameter regime $f = f_c = 1$) \cite{risken}. %It means that in such a limit the amplitudes of $\delta$-kicks are sufficiently large for the particle to overcome the potential barrier.
We emphasize that  colossal enhancement of the relative diffusion coefficient $D/D_T$, as well as the average velocity $\langle v \rangle$, caused by  $\eta(t)$ is not restricted  to the critical tilt regime $\langle \eta(t) \rangle = f_c \sim 1$ but occurs as well for a subcritical regime $f < f_c$ and a supercritical regime $f > f_c$, c.f. Fig. \ref{fig1} as well as  \mbox{Fig. \ref{fig2}.} This phenomenon is particularly pronounced at low temperature regimes where in the system driven by the static tilting force $f$ the crossing events of thermal noise induced escape over the potential barrier are scarce.

There are two characteristic time scales for the dynamics described by Eq. (\ref{model_eta}) that allow to clarify  the colossal enhancement of diffusion, see also Appendix. The first characteristic time is $\tau_0 = \Gamma L^2/\Delta U$ and characterizes  relaxation from a maximum of the potential $U(x)$ to its minimum.   The second characteristic time is $\tau_{\lambda} = 1/\lambda$, i.e.  the inverse of the spiking rate of Poissonian noise $\eta(t)$. In the present study $\tau_0$ is chosen as the characteristic unit of time. Therefore its role can be easily deduced e.g. from Fig. \ref{fig2}. If $\tau_{\lambda} \ll \tau_0$ there is no  colossal enhancement and the diffusion coefficient corresponds to giant diffusion. If $\tau_\lambda \approx \tau_0$ the diffusion is already pronouncedly enhanced over the giant diffusion situation and for $\tau_{\lambda} \gg \tau_0$ one can observe colossal diffusion. Alternatively, if $\lambda \tau_0 \ll 1$, i.e. in the time interval $(0, \tau_0)$ there is  a small number of $\delta$-kicks of large amplitudes then colossal diffusion occurs.

\begin{figure*}[t]
	\centering
	\includegraphics[width=0.45\linewidth]{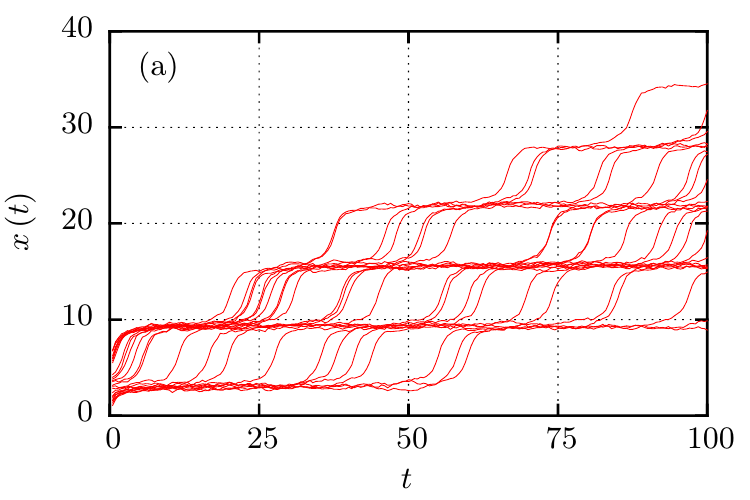}
	\includegraphics[width=0.45\linewidth]{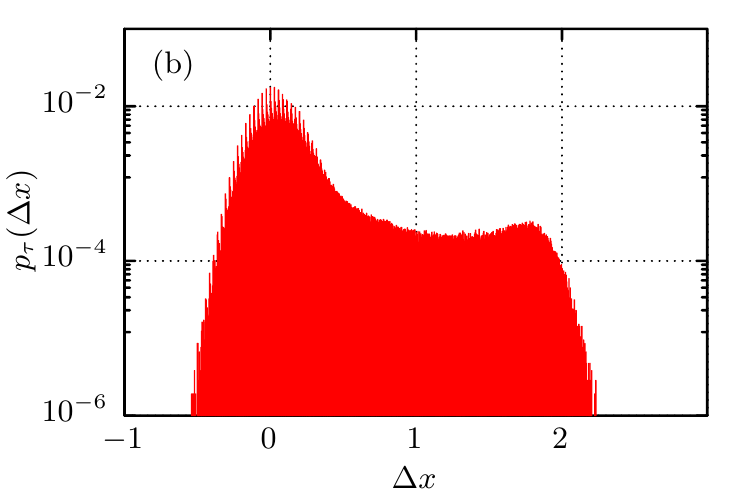}
	\caption{(color online) Overdamped Brownian particle moving in a tilted periodic potential  driven by the static force $f=1$. Panel (a): An exemplary set of realizations of the system trajectories versus elapsed time $t$. Panel (b): The probability density $p_{\tau}(\Delta x)$ for the long time particle position increments  is depicted for the time difference $\tau = 1$.  The thermal noise intensity  $D_T = 0.01$. }
	\label{supp_fig3}
\end{figure*}

In Fig. \ref{fig3} (a) we present a collection of trajectories of the system driven by PSN. We note there the time intervals of relaxation towards the potential minima as well as the long jumps of many spatial periods of the potential caused by $\delta$-spikes. The latter excursions are responsible for such impressive enhancement of diffusion. The corresponding panel for the system under action of the constant bias $f$ is presented in \mbox{Fig. \ref{supp_fig3} (a)}. The particle dynamics depicted there is radically different. The reader can observe two processes: thermal noise induced escape from the potential minimum and relaxation towards the next minimum rather than long excursion which is visible when PSN acts on the particle. We stress that for both scenarios diffusion is asymptotically normal, meaning that the variance of the particle position scales linearly with time, i.e. $\sigma_x^2(t) \sim D t$. Moreover, it is known that for the overdamped Brownian particle moving in a tilted periodic potential the PDF  of the particle coordinate $x$ is Gaussian-like \cite{sivan2018}. In panel (b) of  Fig. \ref{fig3}  we present $P(x,t)$ for the system given by Eq. (\ref{model_eta}). %Similarly, it depicts Gaussian-like profile what may be better quantified by calculation of the excess kurtosis
The Gaussianity of this PDF can be quantified by the kurtosis $K(t)$, reading
\begin{equation}
K(t) = \frac{\left \langle [x(t) - \langle x(t) \rangle]^4 \right \rangle}{\left\{ \left\langle [x(t) - \langle x(t) \rangle]^2 \right \rangle \right\}^2} - 3 \;,
\end{equation}
which for the Gaussian density assumes zero, i.e. \mbox{$K(t) = 0$}. In the studied case $K(t)$ calculated in the asymptotic long time limit $t = 10000$, for which the particle diffusion is already normal, yields approximately zero, i.e. $K(t) \approx 0$ and thus the PDF for the particle coordinate $P(x,t)$ is very close to  Gaussian statistics. %which in this case vanishes $K \approx 0$.

Next, we consider the PDF $p_{\tau}(\Delta x)$ of the particle position increments
\begin{equation}
	\Delta x(\tau) = \lim_{t \to \infty}[x(t + \tau) - x(t)],
\end{equation}
where $\tau$ is the time increment. This quantity differentiates between the dynamics induced by the deterministic force  $f$ and the stochastic bias $\eta(t)$. In Fig. \ref{fig3} (c) and (d) we depict this characteristics for the particle driven by  $\eta(t)$ with $\tau = 0.1$ and $\tau = 1$, respectively. In Fig. \ref{supp_fig3} (b), we present it for the case of the static force $f$ with the time difference  $\tau = 1$. For this case the PDF $p_{\tau}(\Delta x)$ can be well approximated by the sum of two Gaussian densities representing the increments originating from the relaxation of the particle towards the potential minimum as well as thermal noise induced crossing of the potential barrier. In contrast, when PSN acts on the particle then  $p_{\tau}(\Delta x)$ is distinctly non-Gaussian. Moreover, its tail is characteristic for the class of Laplace distributions $p_{\tau}(\Delta x) \sim e^{-\Delta x}$, note the exponential fits (in green) in panels (c) and (d). The impact of the time-lag  $\tau$ on the distribution $p_{\tau}(\Delta x)$ is visualized there as well. Firstly, for increasing  $\tau$ the cut-off of the PDF grows. Secondly, in this latter case the reader can detect the multi-peaked, comb-like shape of the distribution $p_{\tau}(\Delta x)$, which is characteristic for an overdamped dynamics in a periodic potential in which the particle quickly relaxes towards neighbouring potential minima.

\section{Conclusions}
With this study we revealed a new manifestation of Brownian, yet non-Gaussian diffusion. Its characteristic features are that the particle diffusion still proceeds normal with the PDF of the particle position remaining Gaussian-like, the corresponding density for its increments, however, noticeably deviates from the usual Gaussian shape and exhibits an exponential tail. The latter feature results in colossal enhanced diffusion, distinctly {\it surpassing} in magnitude the case of giant diffusion \cite{reimann2002}, obtained upon applying a deterministic bias. In contrast to recent works in the area of Brownian, yet non-Gaussian diffusion this peculiar behaviour is solely a consequence of the external stochastic forcing acting on the particle. This feature opens a new avenue within the recently established and growing activity of non-Gaussian diffusion dynamics in which the nonequilibrium state created by the external perturbations serves as the seed for various kinds of diffusion anomalies \cite{spiechowicz2016scirep,spiechowicz2017scirep}.

In conclusion, we considered a paradigmatic model of nonequilibrium statistical physics consisting of an overdamped Brownian particle diffusing in a periodic potential. This setup comprises numerous experimental realizations \cite{risken,lee2006,reimann2008,ma2015,lutz2013,dechant2019} and therefore we are confident that our findings will inspire and invigorate a vibrant follow-up of both experimental and theoretical studies. Our results has impact as well on a description of biological systems which knowingly operate under nonequilibrium conditions while exposed to non-thermal and non-Gaussian stochastic forces. The result of an exponential tail for position increments leads to a colossal amplification of diffusion coefficient which in addition carries striking consequences on a broad spectrum of the first arrival problems, as e.g. physical and chemical reactions occurring in living cells \cite{lanoiselee2018,epstein2016}.
\section*{Acknowledgement}
This work has been supported by the Grant NCN No. 2017/26/D/ST2/00543 (J. S.).

\appendix*
\section{Dimensionless units}
In physics relations between scales of length, time and energy, but not necessarily  their absolute values play a role  in determining the observed phenomena. Therefore, it is useful to transform the equations describing the model into their dimensionless form. It often allows to simplify the setup description as after such a re-scaling procedure a number of relevant  parameters appearing in the corresponding dimensional version can be reduced.  Moreover, recasting into the dimensionless variables ensures that the obtained results are independent of specific chosen setups, which is essential to facilitate the choice in realizing the best scheme for testing theoretical predictions in experiments. The dimensional versions of the overdamped Langevin dynamics  read
\begin{subequations}
\label{dim_model}
\begin{align}
	\Gamma \dot{x} &=-U'(x)+\sqrt{2\Gamma k_B T}\,\xi(t) + F, \\
	\Gamma \dot{x} &=-U'(x)+\sqrt{2\Gamma k_B T}\,\xi(t) + \eta(t).
\end{align}
\end{subequations}
where the potential is assumed to be in the form
\begin{equation}
	U(x) = \Delta U \sin{\left(2\pi \frac{x}{L} \right)}.
\end{equation}
The parameter $\Gamma$ represents the friction coefficient,  $F$ and $\eta(t)$ stands for the deterministic force and the biased Poisson noise, respectively,  $k_B$ is the Boltzmann constant and $T$ is thermostat temperature. Thermal fluctuations are modelled by $\delta$-correlated Gaussian white noise $\xi(t)$ 
of vanishing mean $\langle \xi(t) \rangle = 0$ and the  correlation function $\langle \xi(t)\xi(s) \rangle = \delta(t-s)$. 
 %with zero mean and unit intensity, i.e.
%\begin{equation}
%	\langle \xi(t) \rangle = 0, \quad \langle \xi(t)\xi(s) \rangle = \delta(t-s).
%\end{equation}
To make Eqs. (\ref{dim_model}) dimensionless we rescale the particle coordinate and time as
\begin{equation}
	\label{scales}
	 \hat{x}=\frac{2\pi}{L}x, \quad \hat{t}=\frac{t}{\tau_0}, \quad \tau_0=\frac{1}{4 \pi^2} \frac{\Gamma L^2}{\Delta U}.
\end{equation}
After such transformations the equations reads
\begin{subequations}
\label{dimless_model}
\begin{align}
	\dot{x} &=-\hat{U}'(\hat{x}) + \sqrt{2D_T}\,\hat{\xi}(\hat{t}) + f , \\
	\dot{x} &=-\hat{U}'(\hat{x})+\sqrt{2 D_T}\,\hat{\xi}(\hat{t}) +\hat{\eta}(\hat{t}),
\end{align}
\end{subequations}
where the rescaled potential
\begin{equation}
	\hat{U}(\hat{x}) = \frac{1}{\Delta U} \, U\left( \frac{L}{2\pi} \hat{x} \right) = \sin{\hat{x}}
\end{equation}
possesses the period $2\pi$ and the barrier height $2$. Other dimensionless parameters are as follows
\begin{equation}
	\label{dimless_params}
	\gamma = 1, \quad f = \frac{1}{2\pi}\frac{L}{\Delta U} F, \quad D_T = \frac{k_B T}{\Delta U}.
\end{equation}
The dimensionless thermal noise takes the form
\begin{equation}
	\hat{\xi}(\hat{t}) = \frac{1}{2\pi} \frac{L}{\Delta U}\, \xi(\tau_0 \hat{t})
\end{equation}
and possesses the same statistical properties as $\xi(t)$, i.e. it is Gaussian stochastic process with the vanishing mean $\langle \hat{\xi}(\hat{t}) \rangle = 0$ and the correlation function $\langle \hat{\xi}(\hat{t})\hat{\xi}(\hat{s}) = \delta(\hat{t} - \hat{s})$. The rescaled biased Poissonian noise reads
\begin{equation}
	\hat{\eta}(\hat{t}) = \frac{1}{2\pi} \frac{L}{\Delta U}\, \eta(\tau_0 \hat{t})
\end{equation}
and is characterized by the following dimensionless parameters
\begin{equation}
	\hat{\lambda} = \tau_0 \lambda, \quad \hat{D}_P = \frac{D_P}{\Gamma \Delta U}.
\end{equation}
It is statistically equivalent to $\eta(t)$, namely, $\langle \hat{\eta}(\hat{t}) \rangle = \sqrt{\hat{\lambda} \hat{D}_P}$ and $\langle \hat{\eta}(\hat{t})\hat{\eta}(\hat{s}) \rangle - \langle \hat{\eta}(\hat{t}) \rangle \langle \hat{\eta}(\hat{s}) \rangle = 2 \hat{D}_P \delta (\hat{t} - \hat{s})$. In the main part of the paper we used only  dimensionless quantities and we therefore the hat notation $\wedge$ 
is omitted.

%%%%%%%%%%%%%%%%%%%%%%%%%%%%%%%%%%%%%%%%%%%%%%%%%%%%%%%%%%%%%%%%%%%%%%%%%%%%%%%%%%%%%%%%%%
\end{document}